%% file: main.tex
\documentclass[conference]{IEEEtran}
\IEEEoverridecommandlockouts

\usepackage{amsmath,amsfonts,graphicx,hyperref,amssymb,enumitem,multirow,makecell,xcolor,mathtools}
\usepackage{cite}
\usepackage{algorithmic}
\usepackage{textcomp}
\usepackage{subcaption}
\usepackage{makecell}
\usepackage{siunitx}
\begin{document}

\title{Color-based Emotion Representation for \\Speech Emotion Recognition}
\author{
    \IEEEauthorblockN{
        Ryotaro Nagase, 
        Ryoichi Takashima, 
        Yoichi Yamashita
    }
    \IEEEauthorblockA{
        College of Information Science and Engineering, Ritsumeikan University \\
        Osaka, Japan\\
        Email: \{rnagase, rtaka\}@fc.ritsumei.ac.jp, yyama@is.ritsumei.ac.jp
    }
}

\maketitle

\begin{abstract}
Speech emotion recognition (SER) has traditionally relied on categorical or dimensional labels. However, this technique is limited in representing both the diversity and interpretability of emotions. To overcome this limitation, we focus on color attributes, such as hue, saturation, and value, to represent emotions as continuous and interpretable scores. We annotated an emotional speech corpus with color attributes via crowdsourcing and analyzed them. Moreover, we built regression models for color attributes in SER using machine learning and deep learning, and explored the multitask learning of color attribute regression and emotion classification. As a result, we demonstrated the relationship between color attributes and emotions in speech, and successfully developed color attribute regression models for SER. We also showed that multitask learning improved the performance of each task.
\end{abstract}


\input{sections/1_intro}
\input{sections/2_annotation}
\input{sections/3_collection_analysis}
\input{sections/4_color_based_SER}
\input{sections/5_conclusion}

\bibliographystyle{IEEEbib}
\bibliography{refs}

\end{document}

%% file: sections/1_intro.tex
\section{Introduction} \label{sec:intro}
Emotions in speech provide crucial cues that support human communication. 
In recent years, social interest has grown in improving the quality of online communication and in addressing customer abuse. 
Along with this trend, speech emotion recognition (SER) has attracted considerable attention. 
The SER technique predicts emotions conveyed through speech. 
It has been applied in various domains, including the development of conversational agents~\cite{app_conversational_agent} and social robots~\cite{app_social_robots}, the analysis of call center conversations~\cite{macary_allosat} and counseling sessions~\cite{Dehua_ICASSP2024}, and the design of e-learning systems~\cite{Li_elearning}. 

In general, conventional methods of SER are usually divided into two types: categorical and dimensional approaches. 
However, these frameworks have limitations in representing both the diversity and interpretability of emotions.
Categorical emotion recognition predicts discrete emotions, such as happiness, anger, and sadness~\cite{ma-etal-2024-emotion2vec,ma24b_interspeech}.
These methods are easy to understand but cannot describe mixed or unclear emotions because they require predefined classes during training and inference.
In contrast, dimensional emotion recognition predicts continuous scores, such as valence, arousal, and dominance~\cite{Vlasenko_ICASSP2024,Sampath_ICASSP2025}.
Although they can capture more subtle differences, their interpretation is not straightforward and often requires domain-specific knowledge.

To address these limitations, we propose a color-based SER framework, which provides both intuitive understanding and quantitative representation.
Color is not only defined by numerical attributes such as hue, saturation, and value, but is also visually intuitive.
Therefore, this framework is expected to be particularly useful in scenarios that require the intuitive visualization of emotions, such as counseling or e-learning.
As an initial step, in this study, we focused on acted emotional speech in Japanese, and we added emotion labels represented as color attributes to an existing dataset that contained only categorical emotion labels.
We analyzed this dataset and demonstrated the relationship between categorical emotions and color attributes, such as hue, saturation, and value.
We also showed that a color-based SER model can be trained with the same model architectures used in conventional categorical SER.
Furthermore, we conducted the multitask learning of color attribute regression and emotion classification, and demonstrated the complementary relationship between categorical emotions and color attributes.
The contributions of this work are as follows:
\begin{itemize}[topsep=2pt,itemsep=1pt,parsep=0pt,partopsep=0pt]
  \item We demonstrate the relationship between emotions in speech and color attributes.
  \item We present the first SER framework that represents emotions with color attributes and directly predicts them from speech.
\end{itemize}

%% file: sections/2_annotation.tex
\begin{figure}[t]
  \centering
  \includegraphics[width=1.0\columnwidth]{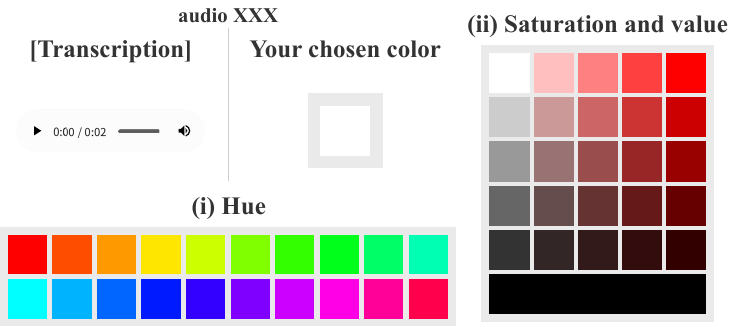}
  \caption{Interface for annotating color attributes}
  \label{fig:ex_hp}
  \vspace{-0.85em}
\end{figure}
\begin{figure*}[t]
  \centering
  \includegraphics[width=1.0\textwidth]{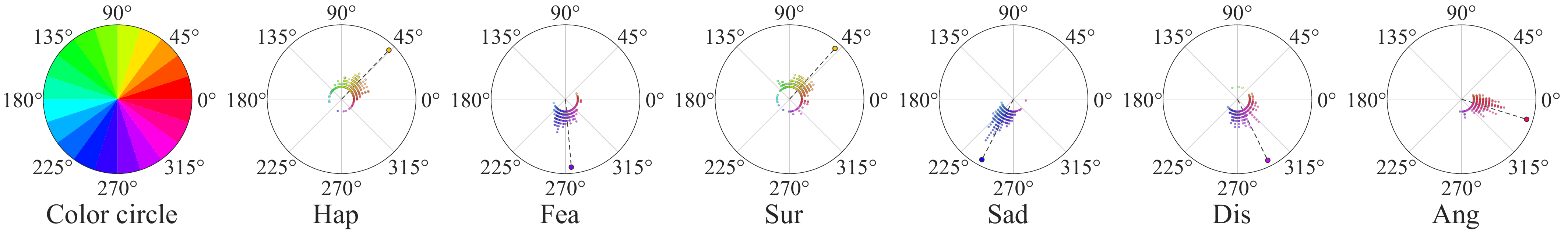}
  \caption{Histogram of hue label frequencies per emotion}
  \label{fig:hue-JVNV}
  \vspace{-0.85em}
\end{figure*}

\section{Procedure for annotating emotions with color attributes} \label{sec:how_to_annotation}
We annotated emotional speech with color attributes, namely hue, saturation, and value. 
Hue represents the type of color ranging from $0^{\circ}$ to $360^{\circ}$, where $0^{\circ}$ corresponds to red, $120^{\circ}$ to green, $240^{\circ}$ to blue, and $360^{\circ}$ to red again.
Saturation represents the vividness of a color ranging from $0\%$ to $100\%$. 
Lower values correspond to grayish colors, whereas higher values indicate more vivid colors. 
Value represents the brightness of a color ranging from $0\%$ to $100\%$, where $0\%$ is black, and with increasing value, the color becomes brighter.
The annotation was conducted through crowdsourcing. 
Annotators used the interface shown in Fig.~\ref{fig:ex_hp}.
They were not given any examples of correspondences between emotional speech and color attributes, and they selected the color on the basis of their own judgment. 
After listening to the speech, they were instructed to ``select the hue that best represents the emotion conveyed by the speech.''
The hue was divided into 20 options at $18^{\circ}$ intervals, presented as tiles shown in Fig.~1(i). 
Then, annotators were instructed to ``select the saturation and value that best represent the emotion conveyed by the speech.''
We provided an interface for them to select saturation and value simultaneously shown in Fig.~1(ii), a design choice based on the findings that these two attributes mutually influence each other.
For example, the Helmholtz--Kohlrausch effect~\cite{Donofrio_HK} indicates that as saturation increases, value is perceived as higher, whereas the Hunt effect~\cite{Hunt_H} indicates that as value increases, saturation is perceived as higher.
Saturation was divided into intervals of $25\%$ up to $100\%$, and value into intervals of $20\%$ up to $100\%$. 
We prepared 26 tile-shaped options, 25 arranged in a $5\times5$ grid, and one additional option representing $0\%$ value.
Annotators could confirm the final choice of color under the label ``Your chosen color.'' 
They were also allowed to modify their selection until they were satisfied.

%% file: sections/3_collection_analysis.tex
\begin{figure*}[t]
  \centering
  \begin{minipage}{0.48\linewidth}
    \centering
    \includegraphics[width=1.0\columnwidth]{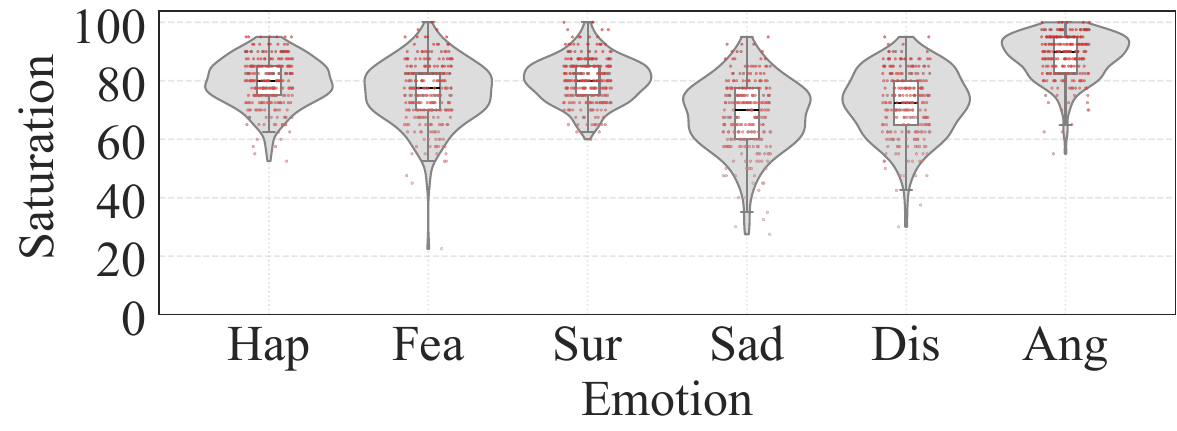}
    \caption{Distribution of saturation label per emotion}
    \label{fig:saturation-JVNV}
    \vspace{-0.85em}
  \end{minipage}
  \hfill
  \begin{minipage}{0.48\linewidth}
    \centering
    \includegraphics[width=1.0\columnwidth]{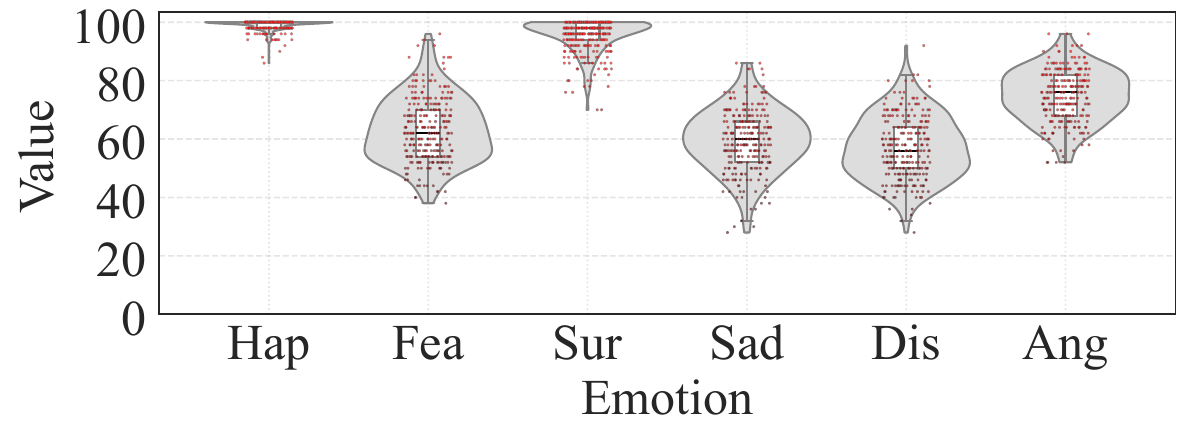}
    \caption{Distribution of value label per emotion}
    \label{fig:value-JVNV}
    \vspace{-0.85em}
  \end{minipage}
\end{figure*}

\section{Collection and analysis of emotions with color attributes} \label{sec:anaysis_color_and_emotion}
\subsection{Emotional speech dataset annotated with color attributes}
In this study, we used the Japanese emotional speech corpus with verbal content and nonverbal vocalizations (JVNV)~\cite{Xin_JVNV}.
This dataset contains Japanese emotional utterances, including nonverbal vocalizations such as laughter, sobs, and screams.
The corpus was recorded by four professional actors in two sessions. 
In the Regular session, they produced verbal and nonverbal utterances designed by the corpus creators, whereas in the Phrase-free session, they produced utterances, including nonverbal vocalizations of their own design.
Each utterance belongs to one of the six categorical emotions, i.e., anger (Ang), disgust (Dis), fear (Fea), happiness (Hap), sadness (Sad), and surprise (Sur).
The corpus contains a total of 1,615 utterances, consisting of 249 anger, 258 disgust, 265 fear, 280 happiness, 257 sadness, and 306 surprise samples.
We collected labels of color attributes from 10 annotators per utterance through the Japanese crowdsourcing platform \textit{Lancers}, following the procedure described in Section~\ref{sec:how_to_annotation}.
After the collection, we averaged hue, saturation, and value and assigned them as the final labels for each utterance.

\subsection{Analysis of collected color attributes with categorical emotions}
\subsubsection{Hue}
Since hue is angular, we evaluated the inter-annotator agreement on hue using the circular standard deviation $\sigma_{\mathrm{circ}}$, as defined in Eq.~\eqref{eq:hue-std}, where $\ln$ is the natural logarithm, $\mathrm{R}$ is the mean resultant length defined in Eq.~\eqref{eq:hue-R}, $\theta_{i}^{\mathrm{rad}}$ is the hue angle in radians of the $i$-th annotator, and $N$ is the number of annotators.
\begin{align}
    \sigma_{\mathrm{circ}} &= \sqrt{-2 \ln \mathrm{R}} \label{eq:hue-std} \\
    \mathrm{R} &= \sqrt{\left(\frac{1}{N}\sum_{i=1}^N \cos \theta_{i}^{\rm rad} \right)^2 + 
              \left(\frac{1}{N}\sum_{i=1}^N \sin \theta_{i}^{\rm rad}\right)^2} \label{eq:hue-R}
\end{align}
The circular standard deviation averaged across utterances was $57.3^\circ$, indicating that annotations tended to cluster within similar hues, such as red-orange or blue-purple.
We also calculated a circular mean as a final label, as defined in Eq.~\eqref{eq:hue-mean}.
\begin{equation}
    \bar{\theta} = \mathrm{arctan2} \left( \frac{1}{N} \sum_{i=1}^{N} \sin \theta_{i}^{\rm rad}, \frac{1}{N} \sum_{i=1}^{N} \cos \theta_{i}^{\rm rad} \right) \label{eq:hue-mean}
\end{equation}

The color circle and circular histogram of hue labels for each emotion are shown in Fig.~\ref{fig:hue-JVNV}.
As the number of utterances assinged the same score increases, the connected line segments become longer.
Each point is colored according to its hue label, and the dashed line indicates the mean hue for each categorical emotion. 
The mean hue is determined to be $46^\circ$ for happy, $275^\circ$ for fear, $48^\circ$ for surprise, $242^\circ$ for sadness, $296^\circ$ for disgust, and $343^\circ$ for anger.

Fig.~\ref{fig:hue-JVNV} illustrates that points cluster at certain angles, with fewer points in the opposite directions across all categorical emotions.
Since JVNV consists of speech performed by professional actors, the hue labels tend to be more consistently matched across utterances.
According to the hue distribution, happiness and surprise cluster around 45$^\circ$, corresponding to yellow, orange, and yellow-green. 
Anger clusters around 0$^\circ$, with a mean hue near 340$^\circ$, corresponding to reddish-purple.
Fear, sadness, and disgust cluster around 270$^\circ$, corresponding to blue and purple, with distinct mean directions.
These results suggest that hue distributions tend to differ across emotions, and that these color attributes may provide additional cues for training an emotion classifier.

\subsubsection{Saturation} \label{sess:saturation}
We evaluated the inter-annotator agreement on saturation using the conventional standard deviation.
The standard deviation averaged across utterances was $21.9\%$, indicating that the annotations were generally consistent.
Fig.~\ref{fig:saturation-JVNV} shows the distribution of saturation labels for each emotion.
Each point is colored according to its saturation label.
From the figure, it can also be observed that the saturation of happiness, surprise and anger, which are considered high-arousal emotions, tends to be higher than about $60\%$.
In contrast, fear, sadness, and disgust, which are considered low-arousal emotions, are widely distributed between $20\%$ and $100\%$. 
These results suggest that the distribution of saturation labels aligns with the arousal axis.

\subsubsection{Value}
We evaluated the inter-annotator agreement on value as in Section~\ref{sess:saturation}.
The standard deviation averaged across utterances was $14.6\%$, indicating that the annotations were generally consistent as well.
Fig.~\ref{fig:value-JVNV} shows the distribution of value labels for each emotion.
Each point is colored according to its value label.
From the figure, it can also be observed that happiness and surprise are concentrated between $80\%$ and $100\%$.
In contrast, value labels of anger are distributed between $50\%$ and $100\%$, and those of fear, sadness, and disgust are distributed between $20\%$ and $100\%$.
These results indicate that positive emotions tend to have higher values, whereas negative emotions tend to have lower values.

All final labels for saturation and value are above $20\%$.
Because color changes are difficult to perceive at low saturation and value, it is possible that annotators selected such low scores less frequently to represent emotions.

%% file: sections/4_color_based_SER.tex
\section{Color attribute regression for SER}
\subsection{Experimental setup}
In this study, we conducted two experiments. 
In Experiment~1, we compared the performance of color attribute regression between support vector regression (SVR) and deep neural network (DNN) models to determine whether color attributes representing emotions can be predicted from speech. 
In Experiment~2, we performed the multitask learning of color attribute regression and categorical emotion classification to analyze the relationship between the two tasks.

\subsubsection{Dataset}
For the training of regression models, we used the JVNV dataset annotated with color attributes. 
We performed leave-one-speaker-out cross-validation. 
The Regular session provided training data, whereas the Phrase-free session was conducted for validation and evaluation without speaker overlap. 
The training data were augmented to five times their original size through speed perturbation with factors ranging from 0.9 to 1.1 in steps of 0.05, following previous SER methods~\cite{Aldeneh_speedaug,Atmaja_dataaug}.

\subsubsection{Models and metrics}
\begin{figure}[t]
    \centering
    \begin{subfigure}{\columnwidth}
        \centering
        \includegraphics[width=\linewidth]{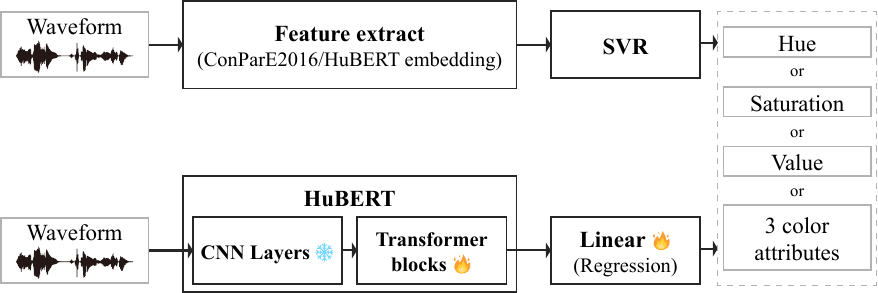}
        \caption{SVR- and DNN-based color attribute regression}
        \vspace{1em}
        \label{sfig:exp1_model}
    \end{subfigure}
    \begin{subfigure}{\columnwidth}
        \centering
        \includegraphics[width=\linewidth]{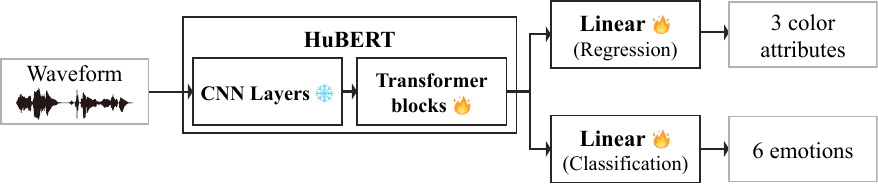}
        \caption{Multitask learning of color attribute regression and categorical emotion recognition}
        \label{sfig:exp2_model}
    \end{subfigure}
    \caption{Outline of the models used in Experiments 1 and 2} 
    \label{fig:model_overview}
    \vspace{-0.85em}
\end{figure}
In Experiment~1, we built two models, which were a SVR model and a DNN shown in Fig. \ref{sfig:exp1_model}. 
The acoustic features used for the SVR model are ComParE2016~\cite{schuller16_interspeech} and embeddings from a Japanese pretrained HuBERT model\footnote{https://huggingface.co/yky-h/japanese-hubert-base} provided by Hugging Face~\cite{wolf-etal-2020-transformers}.
Since previous studies suggested that intermediate layers from the self-supervised learning (SSL) model are effective for SER~\cite{Alexandra_ICASSP2024,Chen_vesper}, we used the outputs of the 6th, 9th, and 12th layers.
When the SSL model outputs were input into the SVR model, the features were converted into fixed-length utterance-level vectors by temporal average pooling. 
The kernel was set to the radial basis function, and other hyperparameters were selected via grid search on the validation data. 
We constructed SVR models independently for each of the color attributes, which are hue, saturation, and value. 
Since hue is an angular variable, we trained separate regressors for sine and cosine components, and reconstructed the hue angle using the $\arctan2$ function.
During DNN training, we combined the pretrained HuBERT model with a regression head consisting of two fully connected layers. 
In the HuBERT model, the parameters of the CNN layers were fixed, and only those of the Transformer blocks were updated, following previous studies on SER~\cite{wang2021fine_tuned_wav2vec_hubert,Wagner_DER}.
We constructed models to predict color attributes individually as well as jointly. 
Note that hue was represented by its sine and cosine components.
The number of epochs was 20, the batch size was 16, and the learning rate was $1 \times 10^{-5}$. 
We used AdamW~\cite{Loshchilov_AdamW} as the optimizer with a linear scheduler.
The loss function was the concordance correlation coefficient (CCC) loss $\mathcal{L}_{\mathrm{CCCL}}$ in Eq.~\eqref{eq:cccl}, where $\rho$ denotes the Pearson correlation coefficient (PCC), $\mu_{y}$ and $\mu_{\hat{y}}$ the means of the ground truth $y$ and the prediction $\hat{y}$, respectively, and $\sigma_{y}^{2}$ and $\sigma_{\hat{y}}^{2}$ their variances.
The CCC loss is widely used in dimensional emotion recognition~\cite{Vlasenko_ICASSP2024,Sampath_ICASSP2025,Wagner_DER}. 
It maximizes the correlation coefficient that accounts for the means and variances of the ground truth and prediction.
\begin{equation}
\mathcal{L}_{\mathrm{CCCL}} = 1 - \frac{2 \rho \sigma_{y} \sigma_{\hat{y}}}{\sigma_{y}^{2} + \sigma_{\hat{y}}^{2} + \left( \mu_{y} - \mu_{\hat{y}} \right)^{2}} \label{eq:cccl}
\end{equation}

In Experiment~2, we used the DNN architecture shown in Fig. \ref{sfig:exp2_model} to perform the multitask learning of color attribute regression and six-categorical-emotion classification. 
The final loss function $\mathcal{L}_{\mathrm{all}}$ is defined in Eq.~\eqref{eq:mtl}, where $\mathcal{L}_{\mathrm{CE}}$ denotes the cross-entropy loss for categorical emotion classification and $\alpha$ is a weighting coefficient. 
$\alpha$ was varied from 0.6 to 1.0 in steps of 0.1.
At $\alpha=1.0$, the model was trained only on categorical emotion classification.
\begin{equation}
\mathcal{L}_{\mathrm{all}} = (1 - \alpha) \mathcal{L}_{\mathrm{CCCL}} + \alpha \mathcal{L}_{\mathrm{CE}} \label{eq:mtl}
\end{equation}

We evaluated hue regression using the angular error (AE) in Eq.~\eqref{eq:maae}, defined as the mean absolute difference between the predicted and ground-truth hues.
\begin{equation}
\mathrm{AE} = \min\!\left(|y - \hat{y}|,\, 360^\circ - |y - \hat{y}|\right) \label{eq:maae}
\end{equation}
We also evaluated saturation and value regression using PCC and CCC, and six-class categorical emotion classification using accuracy.

\subsection{Experiment~1: Comparison of color attribute regression results}
Table~\ref{tab:rslt_svr_and_dnn} shows the performance of hue, saturation, and value regression with the SVR and DNN models.
Note that ``Settings'' correspond to input features for SVR and training methods for DNN.
For the SVR models, HuBERT embeddings, particularly those from intermediate layers, achieved a higher CCC than the traditional feature set. 
In particular, the AE with HuBERT embeddings of the 6th layer improved by $\num{10.4}^\circ$ compared with that with ComParE2016.
These results indicate that SSL features are effective for color attribute regression, which is consistent with previous findings~\cite{Alexandra_ICASSP2024,Chen_vesper}.
For the DNN models, the performance of the individually trained models was slightly higher than that of the jointly trained model.
This result suggests that each color attribute is relatively independent, making simultaneous learning more difficult than single-task learning.
Comparing SVR and DNN models, SVR achieved a lower AE, whereas DNN achieved a higher CCC for saturation and value. 
In addition, the gap between PCC and CCC was smaller for DNN than for SVR. 
These differences are likely due to the fact that CCC was directly optimized.
Overall, the lowest angular error of hue was $\num{31.3}^\circ$. 
This result indicates that predictions remained within similar hue scores rather than across distant ones.
The maximum CCCs were $\num{0.533}$ for saturation and $\num{0.794}$ for value.
This result suggests that both saturation and value can be predicted from speech to some extent.
\begin{table}[t]
  \renewcommand{\arraystretch}{1.3}
  \centering
  \caption{Results of color attribute regression with SVR and DNN models}
  \label{tab:rslt_svr_and_dnn}
  \resizebox{1.0\linewidth}{!}{
    \begin{tabular}{llllll}
    \Xhline{1pt}
    \multirow{2}{*}{Setting} & Hue & \multicolumn{2}{l}{Saturation} & \multicolumn{2}{l}{Value} \\ \cline{2-6} 
    & AE $\downarrow$ & PCC $\uparrow$ & CCC $\uparrow$ & PCC $\uparrow$ & CCC $\uparrow$ \\
    \Xhline{0.75pt}
    \multicolumn{6}{l}{\textit{SVR models}} \\ \Xhline{0.2pt}
    ComParE2016 & \num{41.73943461663144} & \num{0.5637707420091136} & \num{0.32522618486218735} & \num{0.6993273753936798} & \num{0.48005176548979156} \\
    \multicolumn{6}{l}{HuBERT embeddings} \\
    \hspace{1em}$L=6$ & \num{35.14603504} & \num{0.5492906601} & \num{0.3702521997} & \num{0.7547294222} & \num{0.5646740503} \\
    \hspace{1em}$L=9$ & \textbf{\num{31.28580053}} & \num{0.5797376139} & \num{0.4029716581} & \num{0.7875364681} & \num{0.6582187876} \\
    \hspace{1em}$L=12$ & \num{35.1243642} & \num{0.5841542241} & \num{0.3827166625} & \num{0.7291790367} & \num{0.5529699009} \\
    \Xhline{0.75pt}
    \multicolumn{6}{l}{\textit{DNN models}} \\ \Xhline{0.2pt}
    Individual training & \num{34.632802689679764} & \textbf{\num{0.5879133664966745}} & \textbf{\num{0.5329865983319597}} & \textbf{\num{0.8086928070453764}} & \textbf{\num{0.7935932399734184}} \\
    Joint training& \num{35.345425319927536} & \num{0.48342102480451954} & \num{0.4658180904082889} & \num{0.7926840743471117} & \num{0.7714807439376743} \\
    \Xhline{1pt}
    \end{tabular}
  }
  \vspace{-0.85em}
\end{table}

\subsection{Experiment~2: Multitask learning of color attribute regression and emotion classification}
Table~\ref{tab:rslt_mlt} shows the results of multitask learning for color attribute regression and emotion classification with different $\alpha$ values.
As $\alpha$ increased, the CCC of color attribute regression improved. 
Compared with the regression-only model as shown in Table~\ref{tab:rslt_svr_and_dnn}, the multitask learning of regression and classification improved performance by $1.6^\circ$ in hue AE, 0.027 points in saturation CCC, and 0.016 points in value CCC at $\alpha=0.9$.
In addition, the accuracy with multitask learning at $\alpha=0.9$ was 2.5 points higher than that at $\alpha=1.0$.
Specifically, the multitask setting reduced the number of classification errors from sadness to fear and from anger to surprise compared with the classification-only setting. These emotion pairs also differed in mean hue angle.
The results demonstrate that color attribute regression and emotion classification are mutually effective auxiliary tasks.

\begin{table}[t]
  \renewcommand{\arraystretch}{1.3}
  \centering
  \caption{Results of multitask learning of color attribute regression and categorical emotion classification}
  \label{tab:rslt_mlt}
  \resizebox{1.0\linewidth}{!}{
    \begin{tabular}{lllllll}
    \Xhline{1pt}
    \multirow{2}{*}{$\alpha$} & Hue & \multicolumn{2}{l}{Saturation} & \multicolumn{2}{l}{Value} & 
    \multirow{2}{*}{\makecell[l]{Accuracy $\uparrow$\\(6 emotion cls.)}} \\ \cline{2-6}
    & AE $\downarrow$ & PCC $\uparrow$ & CCC $\uparrow$ & PCC $\uparrow$ & CCC $\uparrow$ &  \\ 
    \Xhline{0.75pt}
    0.6 & \num{32.66574139970595} & \num{0.4449901180050824} & \num{0.40616110048398013} & \num{0.775756685051954} & \num{0.7532966909006934} & \num{83.75} \\
    0.7 & \num{31.225794356903343} & \num{0.5001946833987638} & \num{0.4634892684249729} & \num{0.7956074867876218} & \num{0.7838743149728173} & \num{85.83333333333334} \\ 
    0.8 & \num{30.067799593466425} & \num{0.5223671725903785} & \num{0.5038775053256862} & \textbf{\num{0.8214267939158256}} & \textbf{\num{0.8102491975999622}} & \num{88.33333333333334} \\
    0.9 & \textbf{\num{29.736212693404635}} & \textbf{\num{0.5800248569109046}} & \textbf{\num{0.5601425141005889}} & \num{0.8137051432698952} & \num{0.8029119312707225} & \textbf{\num{90.83333333333333}} \\
    1.0 & - & - & - & - & - & \num{88.33333333333333} \\
    \Xhline{1pt}
    \end{tabular}
    }
  \vspace{-0.85em}
\end{table}

\begin{figure*}[t]
    \centering
    \begin{minipage}[t]{0.31\textwidth}
        \centering
        \includegraphics[width=1.0\linewidth]{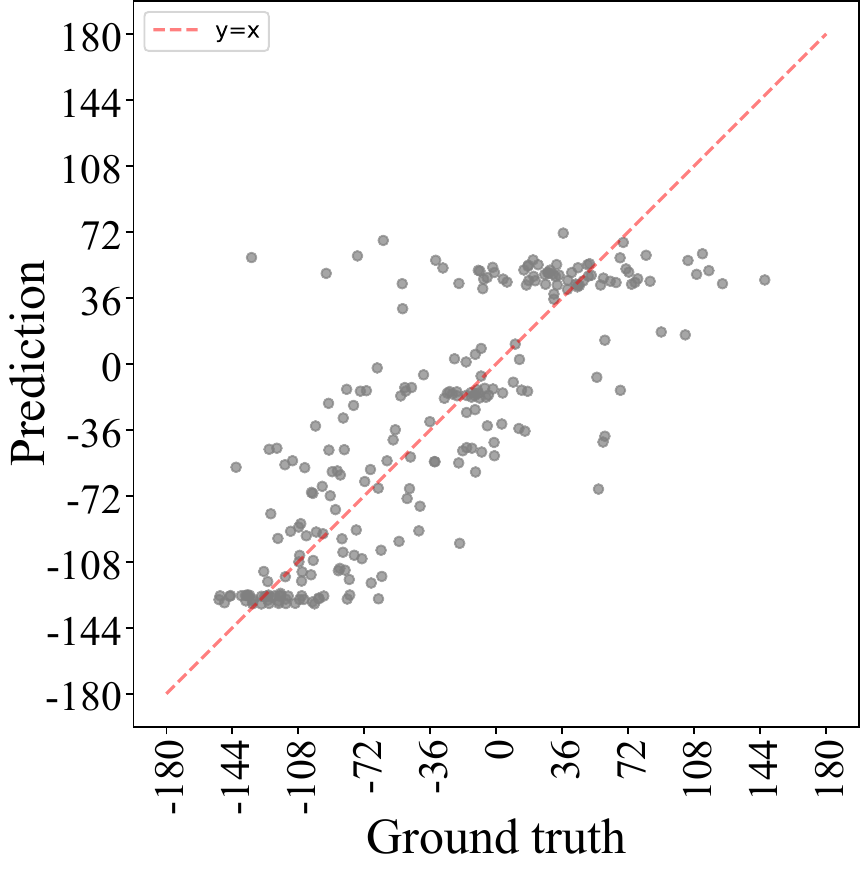}
        \vspace{-1.5em}
        \caption{Prediction result for hue}
        \label{fig:dnn-hue}
    \end{minipage}
    \hspace{0.01\textwidth}
    \begin{minipage}[t]{0.31\textwidth}
        \centering
        \includegraphics[width=1.0\linewidth]{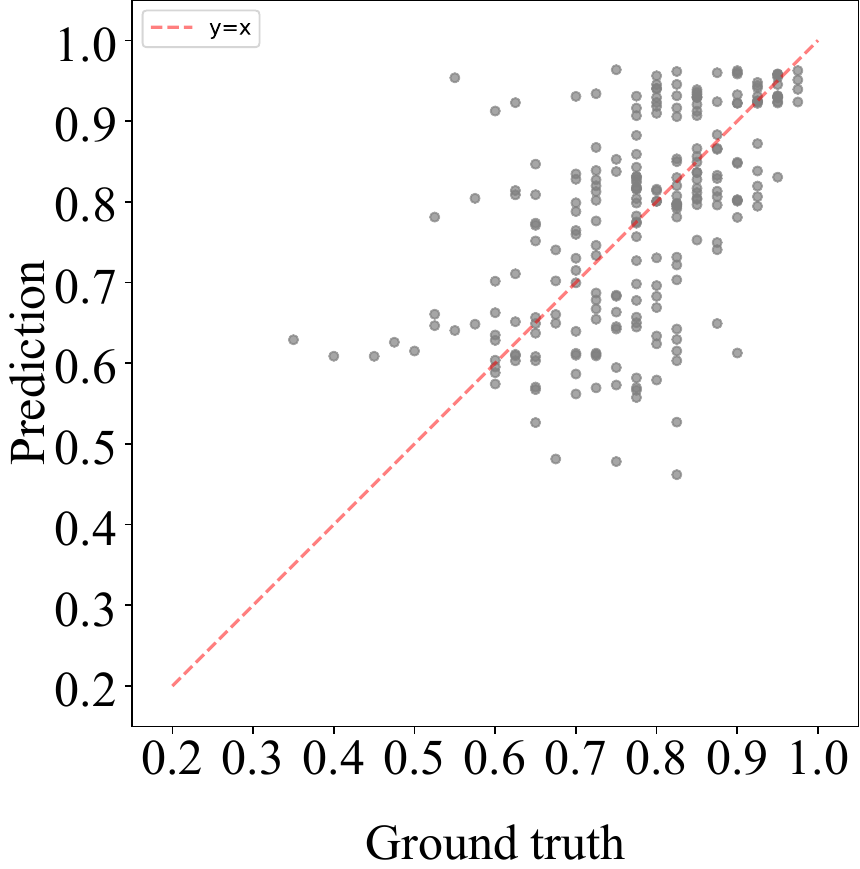}
        \vspace{-1.5em}
        \caption{Prediction result for saturation}
        \label{fig:dnn-saturation}
    \end{minipage}
    \hspace{0.01\textwidth}
    \begin{minipage}[t]{0.31\textwidth}
        \centering
        \includegraphics[width=1.0\linewidth]{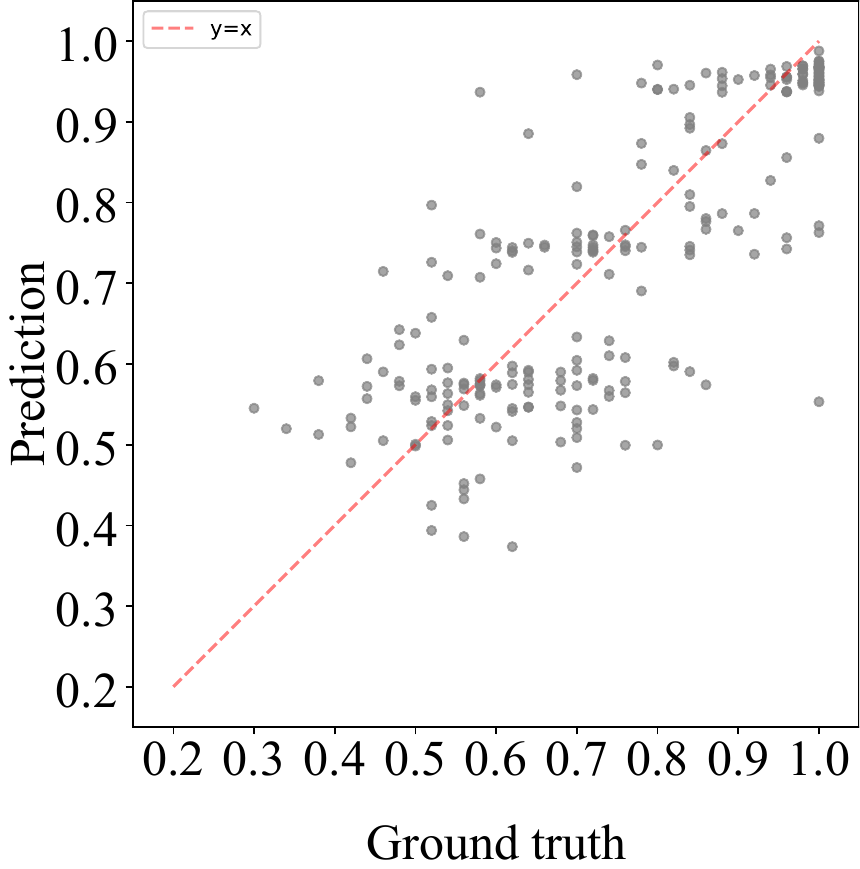}
        \vspace{-1.5em}
        \caption{Prediction result for value}
        \label{fig:dnn-value}
    \end{minipage}
    \vspace{-0.85em}
\end{figure*}
The prediction results for hue, saturation, and value by multitask learning at $\alpha=0.9$ are shown in Figs.~\ref{fig:dnn-hue}--~\ref{fig:dnn-value}.
Note that the horizontal axis is the ground truth score, the vertical axis is the predicted score, and the red dashed line is the ideal score.
The scores of hue are shown in the range of -180$^\circ$ to 180$^\circ$, and those of saturation and value are shown in the range of 0.2 to 1.0.
In Fig.~\ref{fig:dnn-hue}, utterances in happiness and surprise are clustered in the range from $36^\circ$ to $72^\circ$, whereas utterances of sadness and fear are distributed between $-144^\circ$ and $-108^\circ$.
In Fig.~\ref{fig:dnn-saturation}, most predictions are located above 0.5 and are generally consistent with the ideal scores.
In Fig.~\ref{fig:dnn-value}, utterances in fear, sadness, and disgust were distributed between 0.5 and 0.6, those in anger between 0.7 and 0.8, and those in happiness and surprise between 0.9 and 1.0.
These results follow the distributions of hue, saturation, and value labels, and demonstrate that the model trained with the proposed framework can predict each color attribute from speech.

%% file: sections/5_conclusion.tex
\section{Conclusion} \label{sec:conclusion}
In this paper, we proposed a novel color-based emotion representation and clarified its relationship to categorical emotions. 
We also realized a new SER framework that directly predicts emotions as color attributes from speech. 
Moreover, we demonstrate that the multitask learning of color attribute regression and emotion classification has improved the performance in both tasks.
In future work, we will explore the effectiveness of the color-based emotion representation on other types of data, such as spontaneous speech and English speech.